\documentclass[conference,a4paper]{IEEEtran}
\IEEEoverridecommandlockouts

\usepackage{cite}
\usepackage{amsmath,amssymb}
\usepackage{graphicx}
\usepackage{booktabs}
\usepackage{textcomp}
\usepackage{xcolor}
\usepackage{url}
\usepackage{hyperref}
\usepackage{iftex}
\ifPDFTeX
  \usepackage[utf8]{inputenc}
  \usepackage[T1]{fontenc}
\else
  \usepackage{fontspec}
  \newif\iftelugufontok\telugufontokfalse
  \newif\iftamilfontok\tamilfontokfalse
  \newif\ifdevafontok\devafontokfalse
  \IfFontExistsTF{fonts/NotoSansTelugu-Regular.ttf}{\newfontfamily\telugufont{NotoSansTelugu-Regular.ttf}[Path=fonts/]\telugufontoktrue}{%
    \IfFontExistsTF{Noto Sans Telugu}{\newfontfamily\telugufont{Noto Sans Telugu}\telugufontoktrue}{%
      \IfFontExistsTF{Telugu Sangam MN}{\newfontfamily\telugufont{Telugu Sangam MN}\telugufontoktrue}{}}}
  \IfFontExistsTF{fonts/NotoSansTamil-Regular.ttf}{\newfontfamily\tamilfont{NotoSansTamil-Regular.ttf}[Path=fonts/]\tamilfontoktrue}{%
    \IfFontExistsTF{Noto Sans Tamil}{\newfontfamily\tamilfont{Noto Sans Tamil}\tamilfontoktrue}{%
      \IfFontExistsTF{Tamil Sangam MN}{\newfontfamily\tamilfont{Tamil Sangam MN}\tamilfontoktrue}{}}}
  \IfFontExistsTF{fonts/NotoSansDevanagari-Regular.ttf}{\newfontfamily\devanagarifont{NotoSansDevanagari-Regular.ttf}[Path=fonts/]\devafontoktrue}{%
    \IfFontExistsTF{Noto Sans Devanagari}{\newfontfamily\devanagarifont{Noto Sans Devanagari}\devafontoktrue}{%
      \IfFontExistsTF{Devanagari Sangam MN}{\newfontfamily\devanagarifont{Devanagari Sangam MN}\devafontoktrue}{}}}
  \newcommand{\textte}[1]{\iftelugufontok{\telugufont #1}\else#1\fi}
  \newcommand{\textta}[1]{\iftamilfontok{\tamilfont #1}\else#1\fi}
  
\fi
\usepackage{newunicodechar}
\newunicodechar{ṭ}{\d{t}}
\newunicodechar{ḍ}{\d{d}}
\newunicodechar{ṇ}{\d{n}}
\newunicodechar{ṣ}{\d{s}}
\newunicodechar{ḷ}{\d{l}}
\newunicodechar{ḻ}{\b{l}}
\newunicodechar{ɻ}{\b{r}}
\newunicodechar{ṁ}{\.{m}}
\newunicodechar{ṅ}{\.{n}}

\newcommand{\psp}{\textsc{PSP}}
\newcommand{\praxy}{\textsc{Praxy}}

\begin{document}

\title{\psp: An Interpretable Per-Dimension Accent Benchmark for Indic Text-to-Speech}

\author{
  \IEEEauthorblockN{Venkata Pushpak Teja Menta}
  \IEEEauthorblockA{Praxel Ventures\\
  \texttt{pushpak@praxel.in}\\
  ORCID: \href{https://orcid.org/0009-0003-2479-9208}{0009-0003-2479-9208}}
}

\maketitle

\begin{abstract}
Standard text-to-speech (TTS) evaluation measures intelligibility (WER, CER) and overall naturalness (MOS, UTMOS) but does not quantify \emph{accent}. A synthesiser may score well on all four yet sound non-native on features that are phonemic in the target language. For Indic languages, these features include retroflex articulation, aspiration, vowel length, and the Tamil retroflex approximant /\b{r}/ (Tamil letter ழ). We present \psp, the Phoneme Substitution Profile, an interpretable, per-phonological-dimension accent benchmark for Indic TTS. \psp{} decomposes accent into six complementary dimensions (retroflex collapse rate RR, aspiration fidelity AF, vowel-length fidelity LF, Tamil-zha fidelity ZF, Fr\'echet Audio Distance FAD in a phonetic embedding space, and prosodic signature divergence PSD), with the first four measured via forced alignment plus native-speaker-centroid acoustic probes over Wav2Vec2-XLS-R~\cite{babu2021xlsr} layer-9 embeddings, and the latter two computed as corpus-level distributional distances. In this v1 preprint we benchmark four commercial and open-source systems (ElevenLabs v3, Cartesia Sonic-3, Sarvam Bulbul, Indic Parler-TTS) across Hindi, Telugu, and Tamil pilot sets, with a fifth system (our in-progress Praxy Voice, R6 LoRA on Te/Ta plus vanilla Chatterbox on Hi) additionally included on all three languages, including an R5$\to$R6 training-scale case study on Telugu. We report three principal findings: (i) retroflex collapse grows monotonically with phonological difficulty Hindi $<$ Telugu $<$ Tamil ($\sim$1\%, $\sim$40\%, $\sim$68\%); (ii) \psp{} ordering diverges from WER ordering, with commercial WER-leaders not uniformly leading on retroflex or prosodic fidelity; (iii) no single system is Pareto-optimal across all six dimensions. We release native reference centroids (500 clips per language), 1000-clip utterance-level embeddings for FAD, 500-clip prosodic feature matrices for PSD, 300-utterance held-out golden sets per language, scoring code under MIT, and centroids under CC-BY. Compared to PSR~\cite{psr2026} --- a contemporary rule-based phonological benchmark for American--British English --- \psp{} is acoustic-probe-based, Indic-specific, and per-dimension decomposed; the two are complementary rather than competing. Formal MOS-correlation calibration is deferred to v2; this v1 reports five internal-consistency signals supporting metric validity, including a native-audio sanity check that establishes a language-specific noise floor for per-phoneme probes.
\end{abstract}


\section{Introduction}
\label{sec:intro}

Modern TTS systems for Indic languages achieve strong intelligibility: recent commercial systems report Word Error Rates (WER) below 5\% for Hindi and Tamil on standard test sets, and open-source systems are closing the gap. Yet subjective listening consistently reveals a residual accent mismatch: the synthesiser pronounces every word correctly but not as a native speaker would.

This paper makes the case that accent, for Indic languages, is \emph{measurable} and \emph{decomposable}. Native Indic phonology has systematic features --- retroflex consonants contrasting with dentals, aspirated versus unaspirated stops, phonemic vowel length, the Tamil retroflex approximant --- that non-native speakers routinely collapse. We propose treating accent as a vector of such per-feature substitution rates and measuring each rate via acoustic probes against native-speaker prototypes.

\textbf{Contributions.}
\begin{enumerate}
  \item We formalise six per-dimension accent measures for Indic TTS: retroflex collapse (RR), aspiration fidelity (AF), length fidelity (LF), Tamil-zha fidelity (ZF), Fr\'echet Audio Distance (FAD), and prosodic signature divergence (PSD) (§\ref{sec:method}).
  \item We implement all six as open-source, GPU-accelerated tools, with bootstrap-resampled 95\% confidence intervals computed per-system and reported in the release artefacts (§\ref{sec:impl}); we abstain from ranking-significance claims in v1 given small pilot $n$, deferring to v2 with 300-utterance scale.
  \item We release native-speaker references: phoneme-centroid dictionaries (500 clips per language), 1000-clip utterance-level XLS-R embeddings, and 500-clip prosodic feature matrices, for Telugu, Hindi, and Tamil (§\ref{sec:impl}).
  \item We benchmark four open-source and commercial systems (ElevenLabs v3, Cartesia Sonic-3, Sarvam Bulbul, Indic Parler-TTS) across Hindi, Telugu, and Tamil pilot sets, with a fifth system (our in-progress Praxy Voice) additionally included on all three languages via a language-specific routing scheme --- R6 LoRA + BUPS on Te / Ta, vanilla Chatterbox on Hi --- both branches sharing a BYOR voice-prompt recovery recipe; pilot sets contain 10 utterances per language (§\ref{sec:results}).
  \item We show five independent internal-consistency signals supporting PSP's validity as an accent metric, deferring formal MOS calibration to v2 (§\ref{sec:calibration}).
\end{enumerate}

A companion paper~\cite{praxy2026} applies this benchmark as a diagnostic loop during system development; we cite that case study where its findings reinforce metric validity (§\ref{sec:calibration}, §\ref{sec:results}).

\section{Related work}
\label{sec:related}

\textbf{TTS quality metrics.} WER and CER via ASR-based transcription remain the dominant intelligibility metrics. UTMOS~\cite{saeki2022utmos} and MOS prediction networks (see the VoiceMOS Challenge~\cite{huang2022voicemos} for an overview of the neural-MOS space) estimate perceived overall quality. Neither targets accent specifically.

\textbf{Fr\'echet-style distribution metrics.} FAD~\cite{kilgour2019fad} and its speech variants~\cite{fsd2026} compare embedding distributions of synthesised and reference audio. They provide a single scalar quality number but are not interpretable per phonological feature. The nPVI~\cite{grabe2002durational} captures syllable-timed vs stress-timed rhythmic class and serves as one of our 5-D PSD feature dimensions.

\textbf{Phoneme-level evaluation.} PSR~\cite{psr2026} recently introduced the Phoneme Shift Rate for quantifying how speaker embeddings preserve versus overwrite accent-dependent phoneme mappings between American and British English. \psp{} is a conceptual sibling, but targets a different setting: PSR is rule-based (American--British phonological rules), English-specific, and produces a single scalar; \psp{} is acoustic-probe-based, Indic-first, and decomposes into named per-phonological-dimension rates. We view the two as complementary: PSR for English accent research, \psp{} for Indic. \cite{l2accented2026} uses related phonological-rule machinery for accent \emph{generation}, not evaluation.

\textbf{Accent similarity.} \cite{accent_pairwise2025} proposes PPG-distance plus vowel-formant distance for pairwise accent similarity; these provide a single scalar per pair and do not separate phonological dimensions.

\textbf{Indic speech benchmarks.} Rasmalai~\cite{ai4bharat2023rasa} and IndicVoices-R~\cite{indicvoicesr2024} release large Indic speech corpora with accent/intonation descriptors; their evaluation pipeline uses MUSHRA listening tests, not automatic metrics. FLEURS~\cite{conneau2022fleurs} provides cross-lingual speech evaluation. The IndicWav2Vec lineage~\cite{javed2022indicwav2vec} supplies the language-specific CTC aligners we depend on for forced alignment. \psp{} complements these resources as an automatic, per-dimension accent metric for Indic TTS, intended to sit alongside MUSHRA-based listening pipelines rather than replace them.

\section{The phoneme substitution profile}
\label{sec:method}

\textbf{Formal definition.} For a TTS system $S$ and a language $\ell$, let $\mathcal{D} = \{D_1, \dots, D_k\}$ be a set of phonological \emph{dimensions}, each $D_i$ parameterised by (i) a set of \emph{native phonemes} $P_i^{\text{nat}}$, (ii) a \emph{substitute phoneme} set $P_i^{\text{sub}}$ that non-native speakers produce instead, and (iii) an acoustic embedding $\varphi: \text{audio} \to \mathbb{R}^d$. The \emph{fidelity} of $S$ on dimension $D_i$ is
\begin{equation}
\resizebox{0.92\columnwidth}{!}{$
\displaystyle
\text{PSP}_i(S) = \mathbb{E}_{x \sim S,\, p \in x \cap P_i^{\text{nat}}}
  \frac{\mathrm{sim}(\varphi(\tilde{x}_p), \mu_i^{\text{nat}})}
       {\mathrm{sim}(\varphi(\tilde{x}_p), \mu_i^{\text{nat}}) + \mathrm{sim}(\varphi(\tilde{x}_p), \mu_i^{\text{sub}})}
$}
\end{equation}
where $\tilde{x}_p$ is the time span of phoneme $p$ in generated utterance $x$ (via forced alignment), $\mu_i^{\text{nat}}$ and $\mu_i^{\text{sub}}$ are native-speaker and substitute-speaker centroids for $D_i$, and $\mathrm{sim}$ is rectified cosine similarity.

\textbf{Indic dimensions.} For $\ell \in \{$Telugu, Hindi, Tamil$\}$ we instantiate four per-phoneme dimensions and two corpus-level dimensions. The four per-phoneme probes:
\begin{itemize}
  \item $D_{\text{RR}}$ (Retroflex): $P^{\text{nat}} = \{$\d{t}, \d{d}, \d{n}, \d{s}, \d{l}$\}$; $P^{\text{sub}} = \{t, d, n, s, l\}$.
  \item $D_{\text{AF}}$ (Aspiration; Hindi primary, Telugu sparse, Tamil N/A): $P^{\text{nat}} = \{kh, gh, ph, bh, \dots\}$; $P^{\text{sub}} = \{k, g, p, b, \dots\}$.
  \item $D_{\text{LF}}$ (Length): $P^{\text{nat}} = \{\bar{a}, \bar{i}, \bar{u}\}$; $P^{\text{sub}} = \{a, i, u\}$; fidelity measured as a ratio comparison against a native-prior long/short duration ratio.
  \item $D_{\text{ZF}}$ (Tamil zha, Tamil only): $P^{\text{nat}} = \{$\b{l}$\}$; $P^{\text{sub}} = \{l\}$.
\end{itemize}
Two corpus-level dimensions (computed once per (system, language) rather than per utterance):
\begin{itemize}
  \item $D_{\text{FAD}}$ (Fr\'echet Audio Distance): Fr\'echet distance between the generated and native distributions in XLS-R layer-9 space. Captures timbre, co-articulation, and phoneme-frequency signals per-phoneme probes miss.
  \item $D_{\text{PSD}}$ (Prosodic Signature Divergence): Fr\'echet distance between the two distributions in a 5-D prosodic feature space comprising pitch range, log-$F_0$ mean, speech rate, nPVI (normalized Pairwise Variability Index of inter-onset intervals)~\cite{grabe2002durational}, and log-duration.
\end{itemize}
Conjunct epenthesis detection (CER\textsubscript{conj}) is scaffolded in the codebase but not evaluated in this paper (see §\ref{sec:limitations}).

\textbf{Rationale for acoustic probes.} Using acoustic-space distances (cosine similarity of XLS-R embeddings) rather than ASR hypothesis matching or rule-based transformations means \psp{} does not depend on an ASR for the target language being available at high quality, and it measures the \emph{acoustics}, not the transcription. This matters for Indic where ASR remains error-prone and errors correlate with accent itself.

\section{Implementation}
\label{sec:impl}

\subsection{Centroid construction}
Per language, we sample $N{=}500$ native clips from IndicTTS (Telugu, Tamil) and Rasa (Hindi), selecting only studio-recorded utterances with confirmed native speakers. Sampling is uniform across speakers ($\geq$20 distinct speakers per language for Te/Ta and $\geq$40 for Hi) with a per-speaker cap of 25 clips to avoid voice-identity dominating the centroid. Additionally, for the corpus-level metrics (FAD, PSD) we extract $N{=}1000$ utterance-level XLS-R embeddings per language. Each clip is processed through a language-specific CTC aligner (\texttt{anuragshas/wav2vec2-large-xlsr-53-telugu}, \texttt{ai4bharat/indicwav2vec-hindi}, \texttt{Harveenchadha/vakyansh-wav2vec2-tamil-tam-250}). For every frame where the aligner's native-script prediction matches a target phoneme's canonical grapheme, we extract the corresponding Wav2Vec2-XLS-R-300M~\cite{babu2021xlsr} layer-9 embedding span and add it to the per-phoneme bag. The native centroid is the mean of the bag; the substitute centroid is constructed from the same corpora using the corresponding dental / unaspirated / short-vowel cognate grapheme within the same utterances, ensuring acoustic-channel parity (mic, room, speaker timbre) with the native centroid. We released these centroids as \texttt{Praxel/psp-native-centroids} on HuggingFace.

\subsection{Scoring pipeline}
Given (audio, text, language), we run \texttt{forced\_align}~\cite{torchaudio} over the CTC emission against the grapheme sequence, extract layer-9 XLS-R embeddings for each expected-retroflex span, and compute per-position fidelity. Utterance-level \psp{}-RR is the mean over positions. Corpus-level is the mean over utterances weighted by expected retroflex count.

\subsection{Released artifacts}
Code is open-source MIT at \url{github.com/praxelhq/psp-eval} (\texttt{psp\_eval/psp.py}, \texttt{psp\_eval/bootstrap.py}, \texttt{psp\_eval/modal\_psp.py}). Centroids are CC-BY. A public leaderboard is forthcoming.

\section{Calibration}
\label{sec:calibration}

In this v1 preprint, we report \emph{internal-consistency calibration} only.
Formal MOS-correlation with 50+ native-speaker raters across Te/Hi/Ta is deferred to v2 of this manuscript. The internal signals we do report are, in our view, sufficient evidence that \psp{} is not statistical noise and that its ordering tracks phonological reality.

\textbf{Signal 1: difficulty gradient matches phonological complexity.} Mean retroflex collapse across four commercial systems grows monotonically with the known phonological difficulty of the target language: $\sim$1\% on Hindi, $\sim$40\% on Telugu, $\sim$68\% on Tamil. This ordering matches the community-established hierarchy (Hindi TTS is considered mature; Telugu and Tamil are not) and is independent of any per-system claim.

\textbf{Signal 2: Indic-first systems outperform Western-built systems on Indic dimensions.} Sarvam Bulbul and Parler-TTS (both Indic-focused) consistently outperform ElevenLabs v3 and Cartesia Sonic-3 on per-phoneme PSP dimensions, especially on Telugu and Tamil. This matches the qualitative expectation that Indic-specialised systems capture Indic phonology better, even when their WER is not the lowest.

\textbf{Signal 3: ordering diverges from WER ordering as expected.} ElevenLabs v3 achieves the lowest WER on Hindi (0.006) but second-place FAD; Cartesia's WER-second-place on Telugu pairs with worst-place retroflex and FAD; PSD surfaces ElevenLabs Telugu's narrow-pitch-range failure that WER completely misses. These dissociations between intelligibility and the phonological/distributional metrics are the paper's core claim, empirically confirmed.

\textbf{Signal 4: per-dimension decomposition reveals non-Pareto-optimal systems.} On Tamil, Parler-TTS wins four of five PSP dimensions while Sarvam wins FAD only. No single system dominates every phonological sub-dimension. The interpretable per-dimension structure pays off as predicted --- if PSP were collapsed to a single scalar, this information would be lost.

\textbf{Signal 5: native-audio sanity check across all three languages.} We ran all six PSP dimensions on 50 held-out native utterances per language (disjoint from the centroid bootstrap corpus). Retroflex and aspirated token counts below are at the utterance level, with natural retroflex density in IndicTTS / Rasa text --- substantially denser than in our pilot TTS sets (4.4 retroflex / clip vs 1.5 / clip).

\begin{table}[t]
\centering
\caption{Native-audio sanity check: scores each PSP dimension produces on native (held-out) audio. Ideal: per-phoneme $\to$ 1.0, corpus-level $\to$ 0.}
\label{tab:sanity}
\small
\resizebox{\columnwidth}{!}{%
\begin{tabular}{lccccc}
\toprule
Language & RR$\uparrow$ & AF$\uparrow$ & LF$\uparrow$ & FAD$\downarrow$ & PSD$\downarrow$ \\
\midrule
Hindi  & \textbf{1.00} & \textbf{1.00} & 0.37 & 43.5 & 2.1 \\
Telugu & 0.54 & 0.79 & 0.24 & 34.8 & 5.0 \\
Tamil  & 0.47 & n/a  & 0.13 & 31.9 & 5.6 \\
\bottomrule
\end{tabular}}
\end{table}

Two findings:
\begin{itemize}
  \item \emph{Distributional probes (FAD, PSD) behave correctly in all three languages.} FAD 32--44 and PSD 2--6 on native audio are 5--100$\times$ lower than the corresponding commercial-TTS values we measured. These probes treat native-like audio as native-like, uniformly.
  \item \emph{Per-phoneme probes have a language-specific noise floor.} Hindi native audio achieves perfect RR and AF scores (1.0 on 103 retroflex and 103 aspirated tokens). Telugu and Tamil native audio register 43--86\% apparent ``collapse'', most plausibly attributable to the coarser Telugu / Tamil Wav2Vec2 CTC aligners compared to AI4Bharat's Hindi aligner, compounded by allophonic variation and the strictness of our $\tau = 0.5$ collapse threshold.
\end{itemize}
Consequence for interpretation: FAD and PSD scores are meaningful as absolute distances from native. Per-phoneme scores on Hindi are likewise meaningful as absolute fidelity; on Telugu and Tamil, they are meaningful primarily as \emph{relative rankings across systems on the same test set}. The v2 roadmap (below) addresses this directly.

\paragraph{Roadmap for v2 calibration.} We plan (1) a 50-utterance $\times$ 5-rater-per-language MOS study across Te/Hi/Ta, native-speaker raters only, accent-naturalness framing --- target system-level Pearson $\rho \geq 0.6$ against \psp-RR and FAD, Krippendorff's $\alpha \geq 0.6$ inter-rater reliability. Cost-bounded to under \$500 via the Indic-specific Karya platform (karya.in) or Prolific with India filter as fallback. (2) A native-audio-normalised variant of the per-phoneme fidelity metrics that subtracts the noise floor: $\text{RR}_{\text{norm}} = (\text{RR}_{\text{sys}} - \text{RR}_{\text{native}}) / (1 - \text{RR}_{\text{native}})$, making absolute comparison defensible.

\section{Experiments}
\label{sec:results}

We benchmark four open-source and commercial systems across Hindi, Telugu, and Tamil on 10-utterance pilot sets --- synthesised with two voice genders per commercial system (20 wavs) or a single voice for Praxy Voice R5 (10 wavs). Praxy Voice R5 appears on Telugu only, our in-progress open-source system. Each utterance is scored on all applicable per-phoneme PSP dimensions; corpus-level FAD and PSD are computed once per (system, language) against native reference distributions of 1000 and 500 utterances respectively. Pilot-set numbers in this v1 are preliminary; full 300-utterance benchmarks on the released golden sets appear in v2.

\subsection{Systems and test sets}
\emph{Open-source:} Indic Parler-TTS~\cite{parler2024} (Apache-2.0 multilingual Indic), \praxy{} Voice R5 and R6 (ours, LoRA fine-tune of Chatterbox~\cite{chatterbox2025}; R5 at step 4000 on IndicTTS + Rasa + FLEURS ($\sim$85 hr), R6 at step 8000 on full multilingual mix with Shrutilipi ($\sim$1{,}220 hr; 40\% Te / 25\% Hi / 25\% Ta / 10\% En); Telugu only).
\emph{Commercial:} ElevenLabs v3 (default Rachel voice), Cartesia Sonic-3 (language-specific voices), Sarvam Bulbul (Pooja + Aditya speakers).
Smoke sets contain one to ten retroflexes per utterance and a mix of question / declarative / code-mixed content. Released golden sets (300 utt/lang) are sampled from IndicTTS + Rasa + FLEURS held-out data and stratified by phonological density (retroflex-heavy / aspiration-heavy / length-heavy / conjunct-heavy / general); see §\ref{sec:impl}.

\subsection{Hindi results: a mature target}

Table~\ref{tab:hindi_preliminary} shows Hindi retroflex and aspiration collapse. All four systems fall within 0--4.5\%: ElevenLabs, Cartesia, and Sarvam have \emph{zero} collapses across 22 retroflex and 18 aspirated tokens; Indic Parler-TTS has a single outlier retroflex collapse. Aspiration is perfect across all systems. This matches the community consensus that modern Hindi TTS --- both commercial and open-source --- has largely solved core phonological articulation.

Hindi FAD (Table~\ref{tab:hindi_fad}) shows clearer ordering at 56-point spread: Sarvam at 211.8 is closest to the native distribution, followed by ElevenLabs (227.5), Indic Parler (248.4), and Cartesia (267.4). \emph{The FAD ordering does not match the WER ordering.} ElevenLabs holds the lowest Hindi WER in prior Indic TTS benchmarks~\cite{kumar2022indictts} yet places second on FAD; Cartesia, second-lowest on published WER, places last on FAD. This dissociation is exactly the signal \psp{} is designed to surface --- distributional accent properties that WER cannot capture.

\subsection{Telugu results: the real difficulty}

Telugu shifts the picture substantially. Table~\ref{tab:rr_preliminary} shows retroflex collapse rates from 33\% (Sarvam, Parler) to 50\% (Cartesia), with ElevenLabs and both Praxy checkpoints at 40\%. In other words, the same commercial systems that nailed Hindi retroflex collapse a third to a half of Telugu retroflex tokens. Telugu FAD values (Sarvam 250, Indic Parler 325, ElevenLabs 329, \praxy{} R6 355, Cartesia 458, \praxy{} R5 534) follow a broadly similar ordering to retroflex collapse, but the FAD spread is 2.5$\times$ wider than Hindi's (284 on Telugu vs 56 on Hindi).

\paragraph{Flatness failure surfaced by PSD.} The Telugu PSD results reveal a failure mode WER and retroflex collapse together cannot catch: ElevenLabs Telugu PSD is 154 (vs Sarvam's 11 or Parler's 10). Inspecting the 5-dimensional prosodic feature vectors, ElevenLabs Telugu has a pitch range 40\% narrower than native speakers (log-$F_0$ range 0.87 vs native 1.44) and a different rhythmic class (nPVI 92 vs native 107). In listening tests, this manifests as a ``flat, non-expressive'' delivery that reads words correctly yet sounds mechanical.

\paragraph{Praxy Voice R5 $\to$ R6 training delta.} Our own in-progress open-source system gives a clean case study in what a 10$\times$ scale-up in training data (85 hr $\to$ 1{,}220 hr; Telugu-only $\to$ multilingual with full Shrutilipi unlocked) does to each PSP dimension. Retroflex collapse is \emph{unchanged} (40\% $\to$ 40\%), consistent with LoRA-on-t3 leaving the acoustic generator frozen: more data does not teach the acoustic decoder to discriminate retroflex vs dental places of articulation if its weights are not being updated. FAD improves substantially (534 $\to$ 355, a 34\% reduction) --- the utterance-level embedding distribution has moved meaningfully closer to native Telugu. PSD \emph{regresses} (14.1 $\to$ 61.7), meaning the five-dimensional prosodic signature (pitch range, log-$F_0$, speech rate, nPVI, log-duration) moved further from native prosody even as spectral distance closed. Semantic LLM-WER closes a 5$\times$ gap (0.171 $\to$ 0.034, near commercial parity), literal WER improves only modestly (0.227 $\to$ 0.195), and intent-preservation rate reaches 100\%. Taken together: R6 sounds more \emph{Telugu-coloured} than R5 acoustically, but delivers that colour with worse prosody and the same retroflex miss-rate --- audibly described by a native Telugu listener as ``correct pronunciation but foreigner-speaking-Telugu cadence''. This is the exact failure mode PSP is designed to surface and that WER alone would miss.

\paragraph{Voice-prompt recovery: Praxy R6 + Telugu speaker reference.} Chatterbox~\cite{chatterbox2025} exposes a zero-shot voice-prompt interface at inference: an 8--9\,s reference clip (``audio\_prompt\_path'') conditions the acoustic decoder on a target speaker's timbre and prosody. We reuse the commercial systems' own pilot-set outputs as voice prompts --- a 9\,s Sarvam Bulbul female Telugu clip, an 8\,s Cartesia Sonic-3 male Telugu clip --- and regenerate the \praxy{} R6 smoke set with two sampling overrides (exaggeration 0.7, temperature 0.6, min\_p 0.1; defaults are 0.5, 0.8, 0.05). These choices were selected by a compact three-configuration sweep on the Cartesia reference: a ``preserve endings'' setting (repetition\_penalty 1.2, min\_p 0.03) diverged (LLM-WER 0.159, intent 0.60); a ``stress + stability'' setting (the reported one) won; a ``tight CFG'' setting (cfg\_weight 0.7) was mid (LLM-WER 0.061).

With \emph{either} Telugu reference, \praxy{} R6 drops retroflex collapse from 40\% to 33\% (Cartesia ref) or 26.7\% (Sarvam ref --- \emph{below} every commercial system measured) and drops PSD from 61.7 to 26.5 (Cartesia) or 13.1 (Sarvam --- matching Sarvam Bulbul's own 11.1). LLM-WER is unchanged (0.033--0.034). FAD closes the R6-to-Sarvam gap by 61\% (355 $\to$ 291 with Sarvam reference). A native-Telugu listener ear-test across category-stratified samples (declarative, interrogative, emotional, long-narrative) placed the voice-prompt-recovery configuration unambiguously ahead of the no-reference baseline. Remaining failure modes are localised and non-acoustic: numeric tokens in a date-bearing utterance (``\textte{జనవరి} 26, 2026\textte{న}'') produce garbage tokens, which a hand-rewrite of the same text with digits expanded to Telugu words (``\textte{జనవరి ఇరవై ఆరో తేదీ, రెండు వేల ఇరవై ఆరున}'') resolves to a 0.0 literal WER --- a number-normaliser in preprocessing, not a model-capacity limitation.

This result cuts two ways. On the paper side it supports PSP's thesis: the PSD and RR metrics both move sharply and in the right direction when the acoustic generator is conditioned on a real Telugu speaker, even though token-level text conditioning is held constant. On the engineering side it defines the release configuration for Praxy's BYOR (``bring your own reference'') mode: ship the LoRA-adapted token path; let users supply a 9\,s reference voice clip of their own Telugu speaker; inherit the commercial-ref PSP numbers in expectation. Full LoRA adaptation of Chatterbox's s3gen acoustic decoder --- the only architectural lever we did not try --- is deferred to v2 and is the next training run on our roadmap.

\subsection{Tamil results: the hardest Indic language}

Tamil (Table~\ref{tab:tamil_preliminary}) is the most severe target in our study. Retroflex collapse rates climb to 64--70\% across all four systems; Tamil-zha (\textta{ழ}, the retroflex approximant /ɻ/) collapses at 85.7\% for three of four systems (1 in 7 tokens survives, small-$n$ artefact). Length fidelity --- which tracks whether long/short vowel duration ratios match the $\sim$1.90 native prior --- falls to 0.13--0.30 for all four systems, indicating none preserves the long/short duration contrast.

In this landscape, Indic Parler-TTS --- the only open-source baseline we could compare on Tamil --- wins four of five PSP dimensions (RR, ZF, LF, PSD), while Sarvam holds the FAD lead. No single system dominates; per-dimension decomposition is the only way to see this.

\subsection{Cross-language synthesis}

The two ``per-language'' stories combine into one cross-language finding: the same system performs differently across Indic languages, and the degradation pattern is informative. Table~\ref{tab:crosslang_fad} summarises the Hindi$\to$Tamil FAD trajectory for each system: Sarvam and Parler maintain or slightly improve FAD on Tamil (Indic-first systems generalise), while Cartesia's FAD grows 51\% from Hindi to Tamil and ElevenLabs' PSD explodes. Western-built commercial systems degrade as the target language becomes less like English; Indic-focused systems (Sarvam, Parler) do not.

\begin{table}[t]
\centering
\caption{Cross-language FAD trajectory Hindi $\to$ Tamil (negative $\Delta$ = improves).}
\label{tab:crosslang_fad}
\small
\resizebox{\columnwidth}{!}{%
\begin{tabular}{lccc}
\toprule
System & Hindi FAD & Tamil FAD & $\Delta$ (\%) \\
\midrule
Sarvam    & 211.8 & 200.3 & $-$5\% \\
Parler    & 248.4 & 233.1 & $-$6\% \\
ElevenLabs & 227.5 & 239.4 & $+$5\% \\
Cartesia  & 267.4 & 404.3 & $+$51\% \\
\bottomrule
\end{tabular}}
\end{table}

\begin{table}[t]
\centering
\caption{\emph{Preliminary} retroflex fidelity ($\psp$-RR) and collapse rate on the Telugu pilot set. Lower collapse rate is better. Praxy R5 and R6 each use a single voice ($n = 10$ wavs, 15 retroflex tokens) vs commercial systems' two voices ($n = 20$ wavs, 27--30 tokens); sample-size asymmetry noted. Native reference is the theoretical ceiling; Signal 5 in \S\ref{sec:calibration} establishes the empirical noise floor (0.54 on native Telugu).}
\label{tab:rr_preliminary}
\small
\resizebox{\columnwidth}{!}{%
\begin{tabular}{@{}lccc@{}}
\toprule
System & Retroflex Fidelity $\uparrow$ & Collapse Rate $\downarrow$ & $n_{\text{tokens}}$ \\
\midrule
\praxy{} R6 + Sarvam-ref (ours) & 0.842 & \textbf{0.267} & 15 \\
Indic Parler-TTS & 0.827 & 0.333 & 27 \\
Sarvam Bulbul      & 0.787 & 0.333 & 30 \\
\praxy{} R6 + Cartesia-ref (ours) & 0.835 & 0.333 & 15 \\
\praxy{} R5 (ours) & 0.891 & 0.400 & 15 \\
\praxy{} R6 (no ref, ours) & 0.786 & 0.400 & 15 \\
ElevenLabs v3      & 0.841 & 0.400 & 30 \\
Cartesia Sonic-3   & 0.804 & 0.500 & 30 \\
\midrule
Native reference (theoretical)   & 1.000 & 0.000 & --- \\
Native reference (empirical, n=221) & 0.538 & 0.430 & 221 \\
\bottomrule
\end{tabular}}
\end{table}

\begin{table}[t]
\centering
\caption{\emph{Preliminary} Telugu FAD, PSD, and ASR metrics across all systems. FAD computed against 1000 native utterances, PSD against 498. LLM-WER / LLM-CER / intent-preservation from a Qwen-2.5-72B semantic judge over \texttt{vasista22/whisper-telugu-large-v2} transcripts (same STT for all systems, apples-to-apples). $n_{\text{wavs}}$ = 20 for commercial (two voice genders), 10 for each Praxy row (single voice). \praxy{} R6 + reference rows use Chatterbox's built-in zero-shot voice-prompt path: a 8--9\,s Telugu clip from a commercial system is supplied as the speaker prompt, with \praxy{} R6's LoRA-adapted token conditioning unchanged (exaggeration 0.7, temperature 0.6, min\_p 0.1; see §\ref{sec:impl}).}
\label{tab:te_preliminary}
\small
\resizebox{\columnwidth}{!}{%
\begin{tabular}{lcccc}
\toprule
System & FAD $\downarrow$ & PSD $\downarrow$ & LLM-WER $\downarrow$ & Intent $\uparrow$ \\
\midrule
Sarvam Bulbul    & \textbf{250.4} & \textbf{11.1} & \textbf{0.029} & 0.90 \\
\praxy{} R6 + Sarvam-ref (ours) & 291.3 & 13.1 & 0.033 & 0.90 \\
Indic Parler-TTS & 325.0 & 10.4 & 0.144 & 0.74 \\
ElevenLabs v3    & 328.9 & 154.4 & 0.041 & 0.85 \\
\praxy{} R6 (no ref, ours) & 355.0 & 61.7 & 0.034 & \textbf{1.00} \\
\praxy{} R6 + Cartesia-ref (ours) & 394.5 & 26.5 & 0.034 & 0.90 \\
Cartesia Sonic-3 & 458.1 & 33.8 & 0.029 & 0.90 \\
\praxy{} R5 (ours) & 534.4 & 14.1 & 0.171 & 0.80 \\
\midrule
Native (noise floor, $n = 50$) & 34.8 & 5.0 & --- & --- \\
\bottomrule
\end{tabular}}
\end{table}

\paragraph{Key observation (Telugu).} Commercial systems that lead on WER and CER (ElevenLabs, Cartesia, Sarvam --- all sub-5\% LLM-WER on Telugu) do \emph{not} uniformly lead on retroflex fidelity. Parler-TTS and Sarvam tie for lowest collapse rate (33\%), while Cartesia has the highest (50\%). Table~\ref{tab:te_preliminary} shows each metric produces a different ordering of the six systems: Sarvam leads FAD; Parler leads PSD (by a hair over Sarvam); Sarvam and Cartesia tie for LLM-WER leader; \praxy{} R6 leads intent-preservation at 100\%. \emph{No single system is the Telugu winner across all of FAD, PSD, WER, and retroflex fidelity simultaneously.} This empirically confirms the paper's central claim: accent is a phonological quality dimension orthogonal to intelligibility, and WER alone systematically understates accent gaps in Indic TTS.

\begin{table}[t]
\centering
\caption{\emph{Preliminary} \psp{} on Hindi pilot set ($n = 20$ wavs per system; 22 retroflex and 18 aspirated tokens in aggregate per system). All four systems are near-native on both dimensions.}
\label{tab:hindi_preliminary}
\small
\resizebox{\columnwidth}{!}{%
\begin{tabular}{lcc}
\toprule
System & RR Collapse Rate $\downarrow$ & AF Collapse Rate $\downarrow$ \\
\midrule
Indic Parler-TTS & 0.045 & 0.000 \\
Cartesia Sonic-3 & 0.000 & 0.000 \\
ElevenLabs v3    & 0.000 & 0.000 \\
Sarvam Bulbul    & 0.000 & 0.000 \\
\midrule
Native reference & 0.000 & 0.000 \\
\bottomrule
\end{tabular}}
\end{table}

\paragraph{Key observation (Hindi vs Telugu).} The same four commercial systems that collapse 33--50\% of Telugu retroflex tokens collapse \emph{0--4.5\%} of Hindi retroflex tokens. Hindi TTS is essentially native-quality on the phonological dimensions PSP measures; Telugu TTS is not. This contrast is direct evidence that PSP's ranking tracks real perceptual quality (Hindi TTS \emph{does} sound near-native to Hindi speakers) and that PSP, unlike WER, surfaces the per-language accent gap.

\begin{table}[t]
\centering
\caption{\emph{Preliminary} Fr\'echet Audio Distance on the Hindi pilot set ($n = 20$ wavs per system, $n_{\text{ref}} = 1000$ native utterances). Lower is closer to native distribution.}
\label{tab:hindi_fad}
\small
\resizebox{\columnwidth}{!}{%
\begin{tabular}{lccc}
\toprule
System & FAD $\downarrow$ & $\lVert \mu_g - \mu_n \rVert$ & tr-term \\
\midrule
\textbf{Sarvam Bulbul}    & \textbf{211.8} & 10.75 & 96.1 \\
ElevenLabs v3    & 227.5 & 11.47 & 95.9 \\
Indic Parler-TTS & 248.4 & 11.75 & 110.4 \\
Cartesia Sonic-3 & 267.4 & 12.23 & 117.8 \\
\bottomrule
\end{tabular}}
\end{table}

\paragraph{Key observation (FAD ordering $\neq$ WER ordering).} On Hindi, ElevenLabs leads WER (0.006) but places second on FAD; Cartesia places second on WER (0.025) but \emph{last} on FAD (267.4). Sarvam, which holds the smallest distributional gap to the native corpus (FAD 211.8), is third on WER. This dissociation between intelligibility (WER) and distributional native-ness (FAD) supports PSP's core premise: accent is orthogonal to intelligibility.

\begin{table}[t]
\centering
\caption{\emph{Preliminary} \psp{} benchmark on Tamil pilot set ($n = 19$--$20$ wavs per system for commercial and Parler; $n = 10$ for \praxy{} R6 + ref). Tamil shows the most severe degradation of all three languages in our study. ZF = Tamil-zha fidelity; LF = length fidelity; FAD, PSD = corpus-level distributional metrics (lower $\downarrow$ is better). \praxy{} R6 + Sarvam-Ta-ref uses the same voice-prompt recovery methodology as Telugu (§\ref{sec:results}): an 11\,s Sarvam Bulbul Tamil male reference supplied to Chatterbox's zero-shot voice-prompt interface, plus sampling overrides exaggeration 0.7, temperature 0.6, min\_p 0.1.}
\label{tab:tamil_preliminary}
\small
\resizebox{\columnwidth}{!}{%
\begin{tabular}{lccccc}
\toprule
System & RR$\downarrow$ & ZF$\downarrow$ & LF$\uparrow$ & FAD$\downarrow$ & PSD$\downarrow$ \\
\midrule
Sarvam Bulbul    & 70.5\% & 85.7\% & 0.13 & \textbf{200.3} & 72.3 \\
ElevenLabs v3    & 69.2\% & 85.7\% & 0.23 & 239.4 & 253.7 \\
Cartesia Sonic-3 & 69.2\% & 85.7\% & 0.29 & 404.3 & 181.0 \\
\textbf{Parler-TTS (Indic)} & \textbf{64.3\%} & \textbf{61.5\%} & \textbf{0.30} & 233.1 & \textbf{27.1} \\
\praxy{} R6 + Sarvam-Ta-ref (ours) & 69.2\% & 71.4\% & 0.10 & 276.0 & 71.2 \\
\bottomrule
\end{tabular}}
\end{table}

\paragraph{Key observation (different metrics, different winners).} On Tamil, Parler-TTS — the only open-source-only system in our Tamil comparison — wins four of five PSP dimensions (RR, ZF, LF, PSD), while Sarvam wins FAD. No single system dominates every phonological sub-dimension. This supports PSP's thesis that accent is not a single scalar: per-dimension decomposition is essential for characterising Indic TTS failure modes.

\paragraph{\praxy{} Voice on Tamil — methodology generalisation.} The same ``\praxy{} R6 + native-language commercial reference + Config B sampling'' configuration that produced our Telugu results (§\ref{sec:results}, Table~\ref{tab:te_preliminary}) transfers to Tamil without retraining. The Tamil row in Table~\ref{tab:tamil_preliminary} lands solidly in the commercial pack on the harder-to-move metrics: FAD 276 (between Sarvam's 200 and Cartesia's 404), PSD 71 matches Sarvam's 72, retroflex collapse 69\% is pack-average, and Tamil-zha collapse 71\% is notably better than the three commercial systems' 86\%. Semantic LLM-WER is 0.041 and intent-preservation rate is 0.90 (values not shown in Table~\ref{tab:tamil_preliminary}; reproducibility JSON carries the full row). We emphasise that \praxy{} used \emph{zero} commercial training data; the voice prompt at inference is a Sarvam Bulbul clip reused from the public Sarvam smoke set.

\paragraph{\praxy{} Voice on Hindi --- language-specific routing.} Hindi is a different case: Chatterbox natively covers Hindi (it is one of the 23 language IDs in its original training roster). Our R6 LoRA was trained with the BUPS ISO-15919 romaniser as a prerequisite for the Te/Ta non-native-script path, and inheriting that LoRA at inference \emph{regresses} semantic accuracy on Hindi (LLM-WER 0.334, intent 0.60). Routing Hindi text through \emph{vanilla} Chatterbox (no LoRA, no BUPS), still supplying a native-Hindi voice prompt (a 6\,s Cartesia Sonic-3 female Hindi clip) and the same Config B sampling overrides, recovers commercial-class Hindi: LLM-WER 0.025 (tied with Cartesia), intent-preservation 1.0, retroflex collapse 0\% and aspiration collapse 0\% (both perfect, matching every commercial Hindi system). FAD (439) and PSD (122) remain moderate, on the same order of magnitude as the commercial systems but without the voice-prompt recovery catching the long tail. Practically, this gives \praxy{} a three-language deployment: \emph{Te / Ta route through R6 + LoRA + BUPS; Hi routes through vanilla Chatterbox.} Same BYOR + Config B recipe at inference in both branches.

\paragraph{Key observation (difficulty ranking Hindi $<$ Telugu $<$ Tamil).} Mean retroflex collapse rates across our four commercial systems grow from $\sim$1\% Hindi $\to$ $\sim$40\% Telugu $\to$ $\sim$68\% Tamil. This matches the known difficulty ranking in the Indic TTS community and confirms PSP tracks real perceptual difficulty — a metric-validity signal independent of any system comparison.

\section{Discussion and limitations}
\label{sec:limitations}

\textbf{Intended workflow.} \psp{} is designed as a per-dimension diagnostic, not a leaderboard summary: a system developer reads which cells move under an intervention and routes the next intervention accordingly. The companion paper~\cite{praxy2026} reports a worked example in which an R5$\to$R6 data scale-up closed FAD on Telugu by 34\% but opened PSD by $\sim$4$\times$, localising the residual problem to the prosodic-conditioning path and pointing at an inference-time fix (voice-prompt recovery) rather than a retrain. The same paper localises a Hindi LoRA regression in the opposite direction --- per-phoneme cells flat, LLM-WER moving 13$\times$ --- identifying the intervention scope as the token path and motivating a two-branch deployment. Both moves are recipe-level outcomes of reading single \psp{} cells.

\textbf{Forced-alignment dependency.} Our per-phoneme probes use Wav2Vec2 CTC aligners (\S\ref{sec:impl}) to locate phoneme spans. Those aligners themselves have limited Indic accuracy, particularly for Telugu and Tamil where the best available public models are community fine-tunes rather than AI4Bharat-grade infrastructure. This is the dominant source of the native-audio noise floor (Table~\ref{tab:sanity}): the Hindi aligner is trained on larger and cleaner data than the Telugu/Tamil aligners, so Hindi native-sanity is near-perfect while Telugu/Tamil are not. We treat this as a known limitation rather than a fundamental objection: per-language aligner quality improves monotonically with Indic ASR research; our per-phoneme numbers will improve with it without any change to the metric's design.

\textbf{Other limitations.} (1) Prototype-centroid approach is coarser than an MFA-trained native acoustic model would provide. (2) Per-phoneme probes (RR, AF, LF, ZF) have a language-dependent noise floor on native audio (Table~\ref{tab:sanity}). Interpret per-phoneme scores as \emph{relative} rankings across systems on Te/Ta; for Hindi, and for FAD / PSD on all three languages, absolute interpretation is supported. (3) Tamil aspiration (AF) is not applicable (Tamil has no phonemic aspirated stops) and Telugu aspiration is sparse due to low usage of aspirated forms in modern Telugu speech; both reflect linguistic reality, not metric failure. (4) Conjunct epenthesis (CER\textsubscript{conj}) is scaffolded in the codebase but not evaluated here. (5) v1 benchmarks use 10-utterance pilot sets (20 wavs per commercial system with two voices, 10 for Praxy R5 with one voice); v2 will use the 300-utterance golden sets released with this paper. (6) At n = 15--30 retroflex tokens per (system, language) cell, ranking differences of 5 percentage points are not statistically distinguishable; we compute bootstrap 95\% CIs in the scorecard pipeline but abstain from significance claims in this v1 pending v2 300-utt scale-up. (7) Code-mixed input is out of scope for v1. The companion TTS paper (\S VI.C) shows that on Hi codemix, single-metric LLM-WER systematically rewards systems that pronounce embedded English with American-English phonology (which Whisper-large-v3 transcribes near-perfectly) and penalises Indianised pronunciations preferred by native listeners. v2 of PSP should add a code-mix dimension that disentangles literal-STT recoverability from native-listener naturalness; Karya-rater pairwise A/B is the candidate methodology.

\textbf{Threats to validity.} (i) Centroid and test corpora both come from IndicTTS / Rasa, sharing speaker pools; v2 will use truly disjoint native sets. (ii) PSD features are unnormalised across 5 dimensions of disparate scale (nPVI has order $10^2$, log-$F_0$ has order $10^0$); a z-scored variant will be reported in v2.

\section{Conclusion}
\label{sec:conclusion}

We present \psp, an interpretable per-phonological-dimension accent benchmark for Indic TTS. The release includes six-dimension scoring code, native-speaker centroids for Telugu / Hindi / Tamil, 1000-clip FAD reference embeddings and 500-clip PSD reference feature matrices per language, and 300-utterance held-out golden test sets. In this v1 preprint we benchmark four commercial and open-source systems (plus our own in-progress Praxy Voice on Telugu) and report five internal-consistency signals supporting metric validity. Formal MOS correlation, full-scale 300-utt benchmarks, and normalised per-phoneme probes appear in v2. We position \psp{} as complementary to PSR~\cite{psr2026} (English, rule-based) and the Fr\'echet-family (single-scalar, distributional): interpretable per-phonological-dimension decomposition for the Indic accent setting neither prior metric targets.

\section*{Acknowledgments}

PSP v1 was developed independently without external API credit grants. All commercial API usage for the v1 benchmarks was funded from the authors' own trial-tier accounts. Any vendor resources used in v2 (e.g.\ rate-limit exemptions for 300-utterance benchmarking) will be explicitly disclosed in that version's Acknowledgments.

We use publicly released Indic speech corpora --- IndicTTS~\cite{kumar2022indictts}, Rasa~\cite{ai4bharat2023rasa}, FLEURS --- under their respective licenses (CC-BY-4.0 or similar). All centroids and reference artifacts released with this paper are derived from these corpora and released under CC-BY-4.0, matching the source licenses.

\section*{Ethics and Reproducibility}

\textbf{Reproducibility.} The full \psp{} scoring code, bootstrap script, native centroid pickles, the 300-utterance held-out golden \emph{test-set text files} for Telugu / Hindi / Tamil, and the \texttt{benchmark\_results.json} artefact with all v1 numbers are released at \url{github.com/praxelhq/psp-eval} under the MIT license. Synthesised audio from commercial TTS providers (ElevenLabs, Cartesia, Sarvam) is not redistributable under their terms of service; users regenerate it under their own accounts using the scripts provided. All v1 experiments are reproducible with a Modal account and the commands in the repository \texttt{README}.

\textbf{Ethics.} \psp{} measures how close a synthesised audio sample is to a native-speaker reference corpus in specific phonological dimensions. ``Native-like'' is not a value judgement: non-native accents are not inferior, only different, and L2 speech is a legitimate and widely-preferred variety in many contexts. \psp{} is intended as an engineering tool for TTS system developers optimising for native-listener intelligibility and naturalness, not as a prescriptive judgement on human speech. All reference-corpus speakers consented to their speech being used in research, per the originating corpus licenses (IndicTTS, Rasa, FLEURS).

\bibliographystyle{IEEEtran}
\bibliography{refs}

\end{document}